\documentclass[amssymb,aps,prb,showpacs]{revtex4-1}
\include{ams}
\usepackage{amsmath,amssymb,amsthm}
\usepackage{graphicx}
\begin{document}
\title{Enhancement of thermoelectric figure of merit in zigzag graphene nanoribbons with periodic edge vacancies}
\author{D.V. Kolesnikov}\email{kolesnik@theor.jinr.ru}\author{O.G. Sadykova}\email{issaeva@theor.jinr.ru}\author{V.A. Osipov}\email{osipov@theor.jinr.ru}
\affiliation{Bogoliubov Laboratory of Theoretical Physics, Joint Institute for Nuclear Research,\\ 141980 Dubna, Moscow region, Russia}

\begin{abstract}
 The influence of periodic edge vacancies and antidot arrays on the thermoelectric properties of zigzag graphene nanoribbons is investigated. Using the Green's function method, the tight-binding approximation for the electron Hamiltonian and the 4th nearest neighbor approximation for the phonon dynamical matrix, we calculate the Seebeck coefficient and the thermoelectric figure of merit. It is found that, at a certain periodic arrangement of vacancies on both edges of zigzag nanoribbon, a finite band gap opens and almost twofold degenerate energy levels appear. As a result, a marked increase in the Seebeck coefficient takes place. It is shown that an additional enhancement of the thermoelectric figure of merit can be achieved by a combination of periodic edge defects with an antidot array.
\end{abstract}
\pacs{72.15.Jf, 65.80.-n, 73.63.-b}
\date{\today}
\maketitle
\section{Introduction}

Nowadays, graphene and various graphene-based materials are of considerable theoretical and practical interest.  The increasing development of new experimental techniques, which allow for the controlled growth of graphene nanostructures and manipulation of the graphene edges at the atomic level, has opened up new possibilities for promising applications. Among them is the design of new high-efficiency thermoelectric materials and devices.

The thermoelectric properties of materials are determined by the Seebeck coefficient (S) and the figure of merit (ZT) defined as
$
ZT = {G_e S^2 T}/{\kappa} ,\label{ZTdef}
$
where $G_e$ is the electrical conductance, $\kappa$ is the thermal conductance consisting of the sum of the electronic $\kappa_e$ and phononic $\kappa_{ph}$ contributions, and T is the temperature. Maximum efficiency of a thermoelectric device $\eta_{max}$ can be expressed in terms of ZT and the high/low temperatures $T_H$ and $T_L$ as 
\begin{equation}
\eta_{max} = \frac{T_H-T_L}{T_H}(1-\frac{1+T_L/T_H}{\sqrt{1+ZT}+T_L/T_H}).
\end{equation}
Accordingly, increasing the value of ZT one can increase the efficiency of the device. This determines the overall strategy to increase the thermoelectric efficiency by increasing the Seebeck coefficient and the electrical conductance while reducing the thermal conductance.  

Perspectively, graphene could be considered as a good candidate for a high-quality thermoelectric material due to its very high electronic conductivity~\cite{8k}. However, the practical realization of this possibility faces challenges for two reasons: being zero-gap material graphene has a poor Seebeck coefficient \cite{10k} and, in addition, it demonstrates a very high thermal conductivity~\cite{balandin}. Therefore, in order to achieve high thermoelectric efficiency of graphene one needs to significantly increase the Seebeck coefficient and simultaneously reduce the phonon thermal conductance. 
Many attempts were made to realize this program in various graphene-based structures including nanostructuring in periodic graphene patterns, which results in the bandgap opening~\cite{antidot,sevinci,nanotech}, and incorporation and dimensional tuning of nanopores~\cite{natcomm,sinr}. 

The subject of our interest is the thermoelectric properties of zigzag graphene nanoribbons (ZGNRs). The pristine ZGNR shows high electrical and thermal conductance and negligible Seebeck coefficient. In order to realize good thermoelectric performance in ZGNRs there were considered two possible types of engineering: (i) edge disorder, which drastically reduces phonon thermal transport~\cite{cunni}, and (ii) a combination of extended line defects with impurities and edge roughness, which results in large values of the thermoelectric figure of merit~\cite{kara}. These studies show a possible way how to design a thermoelectric device with high Seebeck coefficient and large values of ZT. For example, in~\cite{kara} a series of design steps were employed. First, introducing extended line defects provides a sharp band edge around the Fermi level and offers a band asymmetry. Additional introducing background impurities allows one to enhance markedly the electronic part of ZT and S. As a third step, introducing edge roughness reduces the lattice part of the thermal conductance. It should be noted that the edge disorder can lead to the increase of ZT by two or three times~\cite{kara,prb91} while periodic nanostructuring enhances ZT up to 3.25 at high temperatures~\cite{sevinci}. On the contrary, poorly controlled random modifications of ZGNRs can reduce the asymmetry of transmission function and, correspondingly, the electronic part of ZT. 

In this paper, we focus on the theoretical design procedure without using random distortions. We suggest to modify edges of narrow ZGNRs by incorporating periodically distributed vacancies and, additionally, introducing  periodic array of holes (antidot lattice). More specifically, we find such type of edge modification in the periodic structure which leads to the step-like behavior of electron transmission function with increased step height due to the coexistence of edge and spatial modes in the electronic spectrum. Moreover, our study shows that the sharp step-like transmission function, which corresponds to the nearly-degenerate conduction or valence band edge states has the best impact on the figure of merit in the periodic structure. This differs from the case of a single pore when both the smooth and sharp transmission functions were found to lead to the increase of ZT~\cite{natcomm}.  We should point out that our emphasis in this paper is in a sense opposite that of~\cite{kara}: the edge reconstruction acts as the source of asymmetry in the electron transmission function while the periodic antidot lattice enhances this effect by a decrease of the lattice thermal conductance thereby increasing the overall impact on ZT.

The paper is organized as follows. In section 2, the methods for the calculation of the Seebeck coefficient, total thermal conductance and ZT are described. We discuss the importance of the phonon and electron transmission functions to determine the value of ZT as well as the influence of defects and antidot arrays on the phonon and electron transmission functions. In section 3, the electron transmission is calculated for the ZGNRs with vacancies symmetrically situated at the edges. We discuss the role of the almost twofold degenerate states appearing, at a certain period of vacancies, at the conduction band edge, as well as the influence of next-nearest-neighbor interaction on these states. Section 4 is devoted to conclusions.

\section{General formalism}

We consider the periodic ZGNR which can be arbitrarily divided into three parts: semi-infinite periodic contacts (left and right) and the finite central part. The electron current and electron thermal current through the device are defined as
\begin{eqnarray}
I_E = \frac{2q}{h}\int_{-\infty}^{\infty} T_e(E)(f(E, \mu_L)-f(E, \mu_R)) dE,\label{Ie}\\
I_Q = \frac{2q}{h}\int_{-\infty}^{\infty} T_e(E)(E-\mu)(f(E, \mu_L)-f(E, \mu_R)) dE,\label{Iq}
\end{eqnarray}
where $T_e(E)$ is the energy-dependent electron transmission function, $f(E,\mu)$ is a Fermi-Dirac distribution function of left and right contacts with chemical potentials $\mu_{L}$ and $\mu_{R}$ , q is the carrier charge, and h is the Planck constant. Assuming the linear response regime one obtains  
from Eqs. (2) and (3) 
\begin{eqnarray}
G_e = q^2 L_0,\;
S = \frac{L_1}{qT L_0},\;
\kappa_e = \frac{1}{T}(L_2-\frac{L_1^2}{L_0}),\label{gpisdefs}
\end{eqnarray}
where
\begin{equation}
L_n(\mu,T) =  \frac{2}{h} \int_{-\infty}^{\infty} T_e(E)(E-\mu)^n (-\frac{df(E,\mu)}{dE})dE.\label{Ln}
\end{equation}
The electron transmission function can be calculated by using the non-equilibrium Green's function formalism (NEGF)~\cite{datta} 
\begin{eqnarray}
T_e(E) = \mathrm{Tr}[\Gamma_L G(E) \Gamma_R G^+(E)], \label{negf0} \\
G(E) = [I(E+i\eta) + H_C-\Sigma_L-\Sigma_R]^{-1},\\
\Gamma_{\alpha} = -2\,\mathrm{Im}\,\Sigma_{\alpha}(E),\; \alpha=L,R,\\
\Sigma_{\alpha} = H_{C\alpha}g_{\alpha}H_{\alpha C},
\label{negf}
\end{eqnarray}
where G(E) is the retarded Green's function of the central part, $\Gamma_{L,R}$ and $\Sigma_{L,R}$ are the broadening function and the self-energy matrix of the left(right) contact, $H_{C}$, $H_{CL}$ and $H_{RC}$ are the Hamiltonian of the central part and blocks corresponding to the interaction between the left(right) contact and the central part, respectively, $I$ is the unit matrix, $\eta$ is a small parameter, and $g_{L,R}$ is the retarded surface Green's function of the left(right) contact. This function is calculated using the Sancho-Rubio scheme~\cite{LopezSancho}. The Hamiltonian matrix is found within the tight-binding approximation.

The phonon thermal conductance can be expressed as 
\begin{equation}
\kappa_{ph} = \frac{1}{2\pi}\int_0^{\infty} T_{ph}(\omega)\hbar\omega\frac{\partial n(\omega,T)}{\partial \omega}d \omega,\label{kph}
\end{equation}
where $T_{ph}(\omega)$ is the phonon transmission function, $\omega$ is the phonon frequency, $\hbar$ is the modified Planck constant, and $n(\omega,T)$ is the Bose-Einstein distribution function. The phonon transmission is calculated using the similar to Eqs. (\ref{negf0})-(\ref{negf}) Green's function approach where the substitution $H\rightarrow K,\;E+i\eta\rightarrow (\omega+i\eta)^2$ should be made, so that the Hamiltonian matrix is replaced with the force constant matrix K (see \cite{Jiang} for detail). We calculate the force constant matrix of the structure within the 4th nearest neighbor approximation by using a set of constants from~\cite{Zimmerman}. Finally, by using (\ref{gpisdefs}) and (\ref{kph}) one can express ZT as
\begin{equation}
ZT = \frac{L_1^2}{T L_0 (\kappa_{ph}+(L_2-L_1^2/L_0)/T)},\label{ZT-L}
\end{equation}
At a fixed temperature, Eq.(\ref{ZT-L}) allows one to obtain ZT as a function of the chemical potential $\mu$ through $T_e(E)$ and $\kappa_{ph}$. 

As it follows from Eq.(\ref{ZT-L}), in order to increase ZT at a certain value of $\mu$ one has to reach a maximum of $L_1^2/L_0$. This condition, for the case of periodic structures, can be fulfilled when $T_e(E)$ has asymmetric step-like behavior \cite{nanotech}. More precisely, the transmission function must experience a sharp increase at the energy $E_0\approx \mu$ from $T_e(E<E_0)=0$ to $T_e(E>E_0+\Delta E)=T_{max}$ in the range of $\Delta E$. This increase will result in the high electron thermal current and significant electronic part of ZT when the conduction or valence band edge falls in the energy window between the left and right electrodes (according to Eq.(\ref{Iq})). At small biases, the width of this window  is determined primarily by the thermal activation energy while the electron thermal current depends on the number of energy states near the band edge that take part in transport (which corresponds to $T_{max}$) and the difference between the energies of nearest levels on the band edge (which corresponds to $\Delta E$). Our analysis based on Eqs.(\ref{Ln}), (\ref{negf0}) and (\ref{ZT-L}) shows that ZT reaches a maximum when the band gap exceeds the value of two or three $k_B T$ (see also Fig.2 in \cite{dress}) and, simultaneously, $\Delta E$ should not exceed this value. 

This leads to a conclusion that structures with a suitable bandgap and nearly degenerate electronic states closest to edges of the valence or conduction band should have both the superior Seebeck coefficient and  thermoelectric efficiency. We found that ZGNRs with periodically distributed edge vacancies having a period equal to three elementary translations demonstrate the desired behavior. The next possible step   to increase ZT is a decrease of $\kappa_{ph}$. This can be achieved by decreasing the width or by introducing antidot arrays into the structure. In our case, we found from (\ref{kph}) at room temperature that the number of atoms in the unit cell, which governs the total number of conducting phonon channels, has a greater effect on the value of $\kappa_{ph}$ than the exact positions of these defects. Notice that in the antidot case the mechanism of electric thermal current generation is preserved, so that the combination of degenerate states at the band edge with the reduced lattice thermal conductance leads to the increase of thermoelectric efficiency.

\section{Results and discussion}

\begin{figure*}
\includegraphics{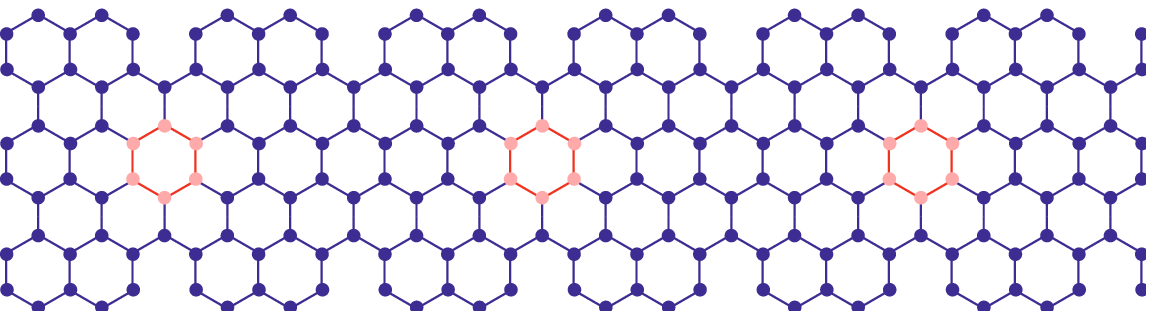} \\
\vspace*{1cm}
\includegraphics{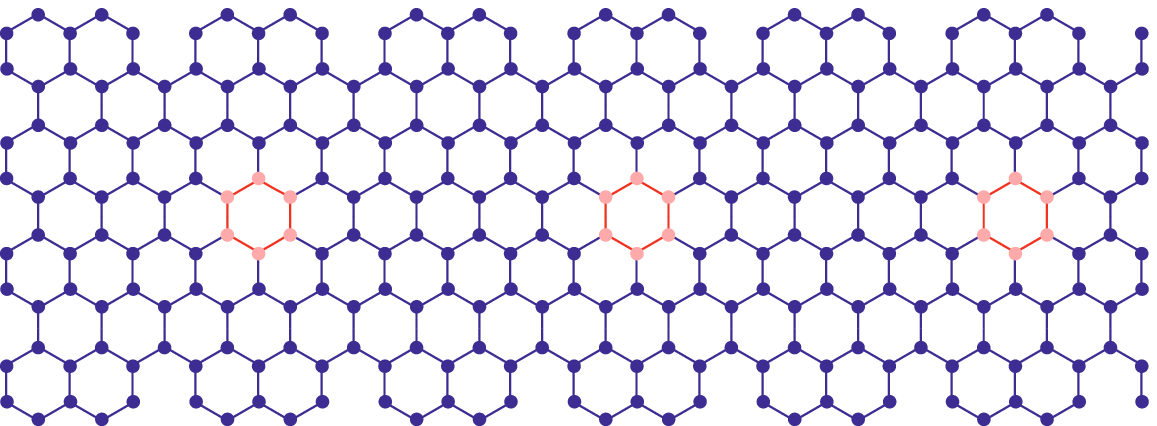}
\caption{ZGNR structures, 6 and 8 atoms wide. The periodic antidot lattice is shown in red.}
\label{fig-latt}
\end{figure*}

\begin{figure*}
\resizebox{0.9\textwidth}{!}{%
\includegraphics{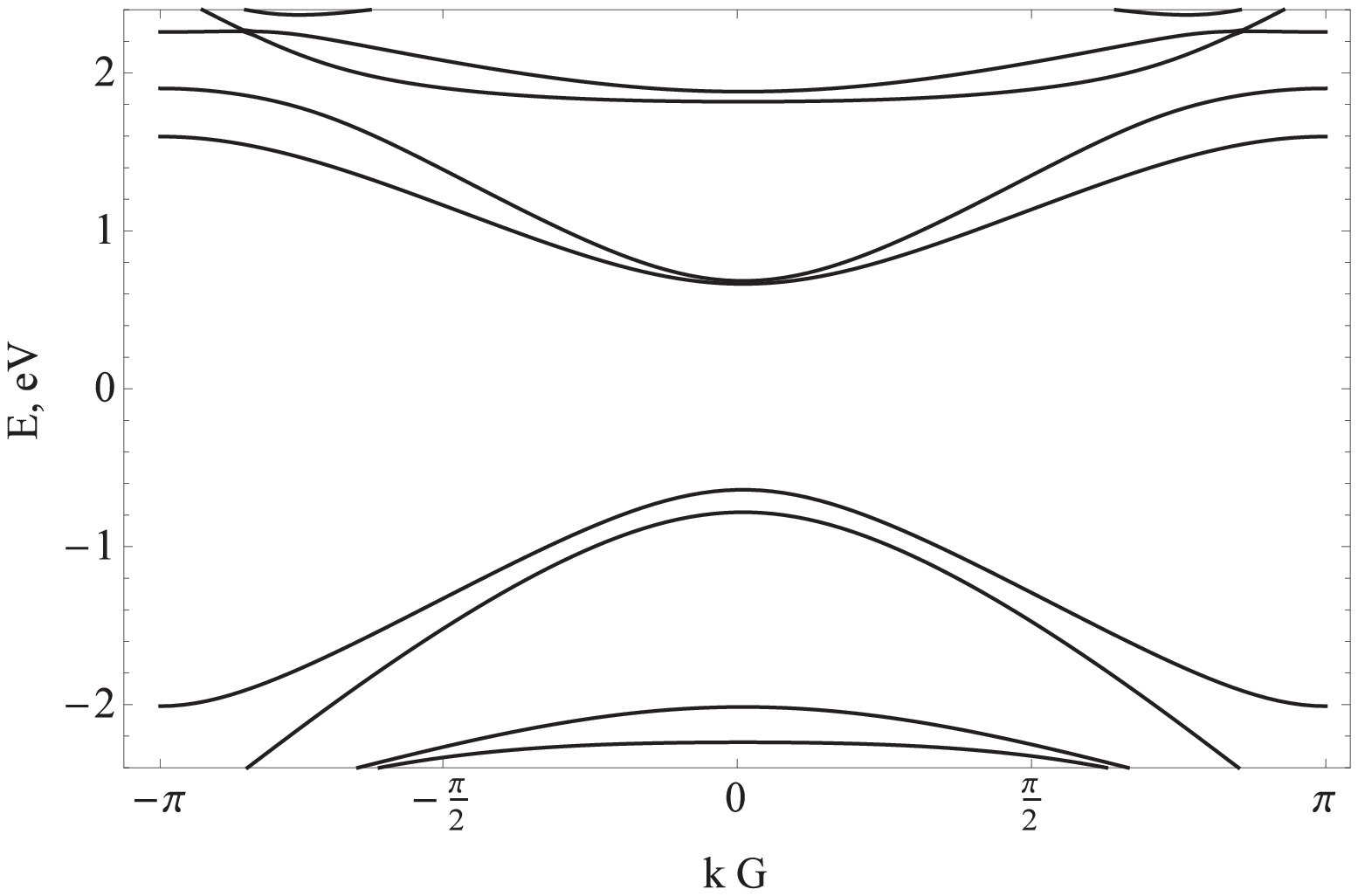} \quad
\includegraphics{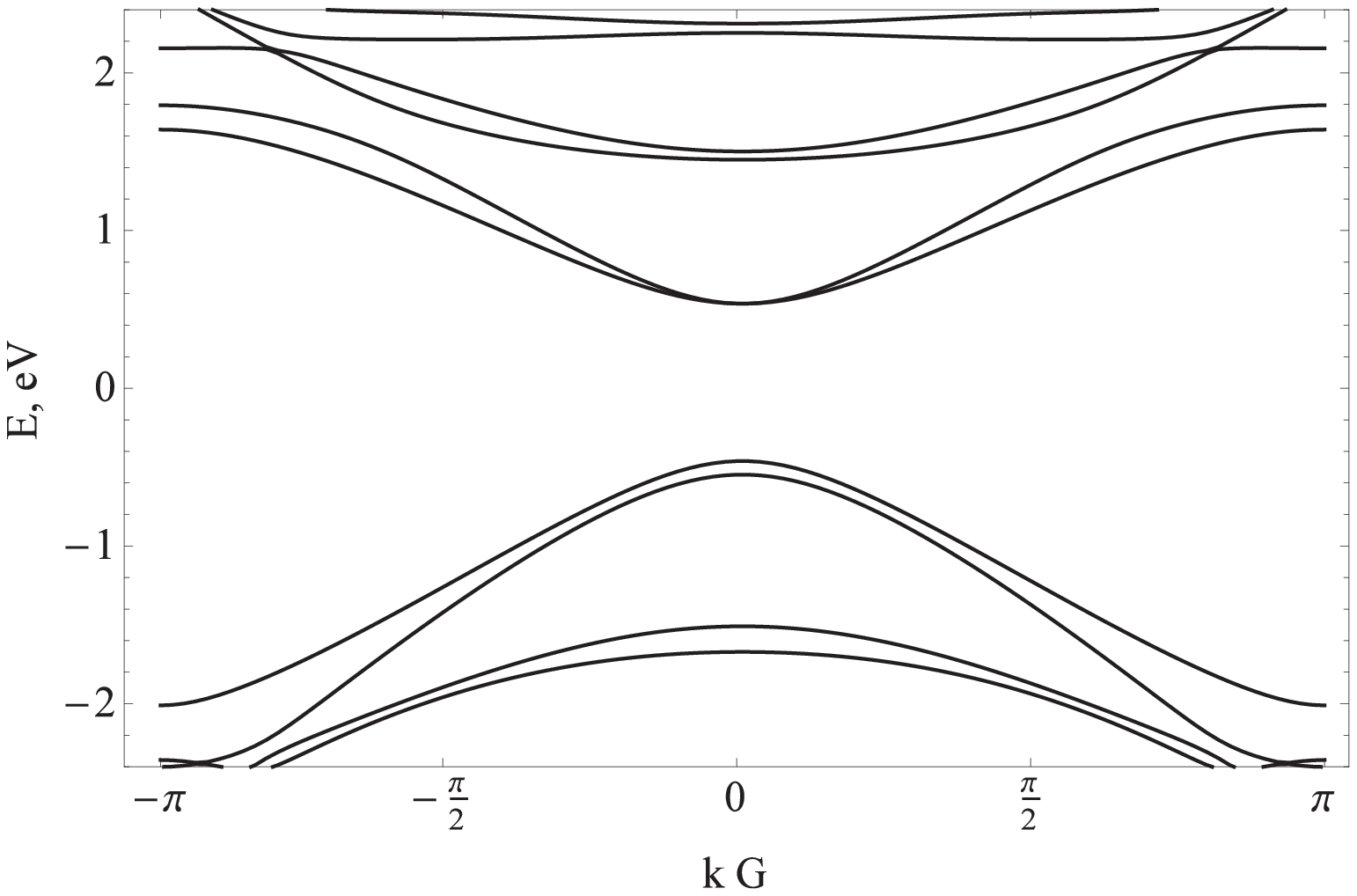} 
}
\caption{Band structures of 6 and 8 atoms wide ZGNRs with edge vacancies: energy (in eV) vs the dimensionless wavevector kG, where G is the spatial period of the structure.
}
\label{fig-band}
\end{figure*}

Fig.~\ref{fig-latt} shows the studied lattice configurations of graphene nanoribbons with periodically arranged edge vacancies and antidots. 
The corresponding electronic band structures are calculated within the tight-biding approximation  with the nearest (NN) and next-nearest-neighbor (NNN) hopping parameters taken as $t=2.7 \;eV$ \cite{prl} and $\;t'=-0.1t$ \cite{Kretinin}, respectively (see Fig.\ref{fig-band}). 
Here and below the energy is measured from the center of the band gap. Notice that there is a shift between NN and NNN cases on the order of 2.5 $t'$ (for comparison, it takes the value of $3t'$ for graphene sheet).
As is seen, the distance between levels $\Delta E$ at the valence band edge is larger than that at the conduction one. Moreover, at the room temperature the states on the edge of the conduction band can be regarded as degenerate. Notice that such behavior is due to the influence of the next-nearest neighbors. Indeed, within the NN approximation the band structure is found to be symmetric so that $\Delta E$ is the same for both bands. 

\begin{figure*}
\resizebox{0.9\textwidth}{!}{%
\includegraphics{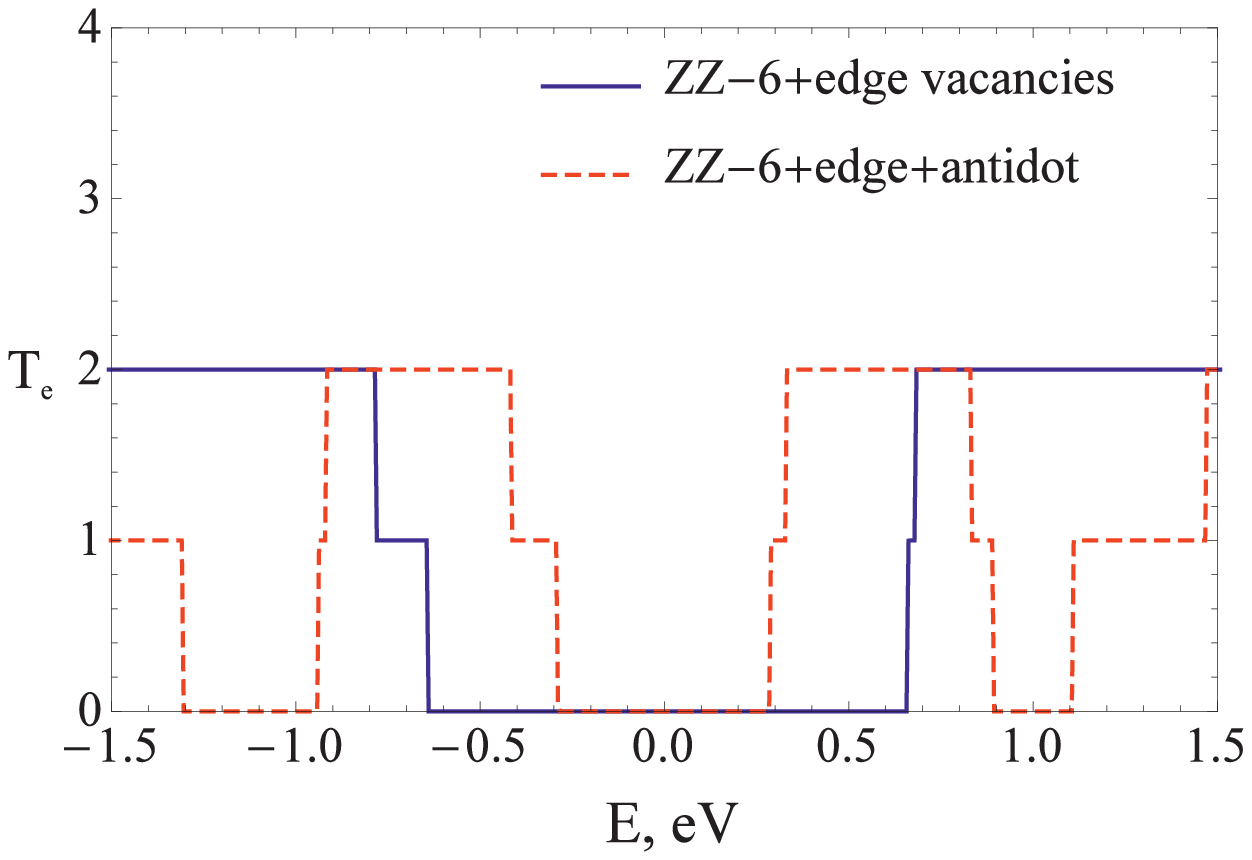} 
\includegraphics{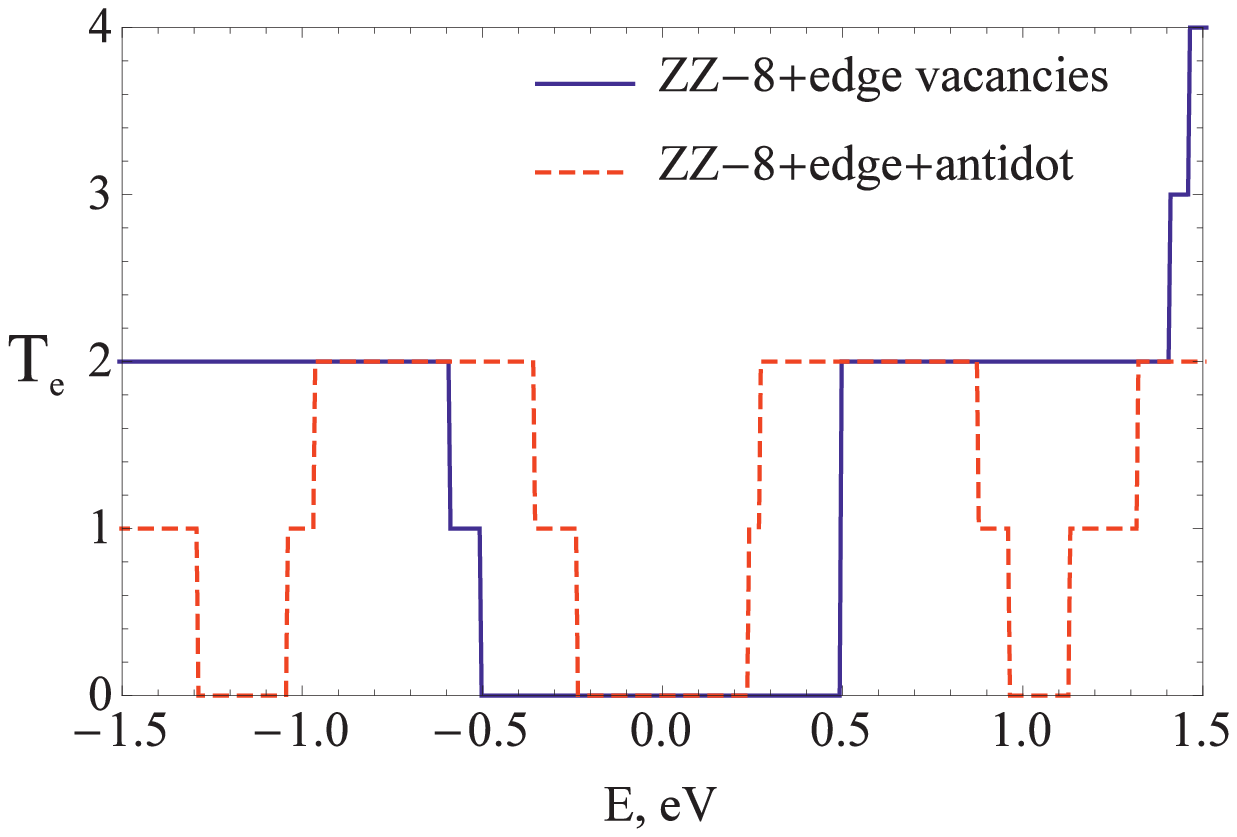}
}
\caption{The electron transmission vs energy for the 6 and 8 atoms wide ZGNR  with edge vacancies (solid line) and additional periodic antidot (dashed line).}
\label{fig-Te}
\end{figure*}

Fig.~\ref{fig-Te} shows the electron transmission functions. 
One can see that the introduction of the periodic antidot lattice results in a decrease of the energy gap, a slight increase of $\Delta E$ and an appearance of additional steps at higher energies.   
\begin{figure*}
\resizebox{0.9\textwidth}{!}{%
\includegraphics{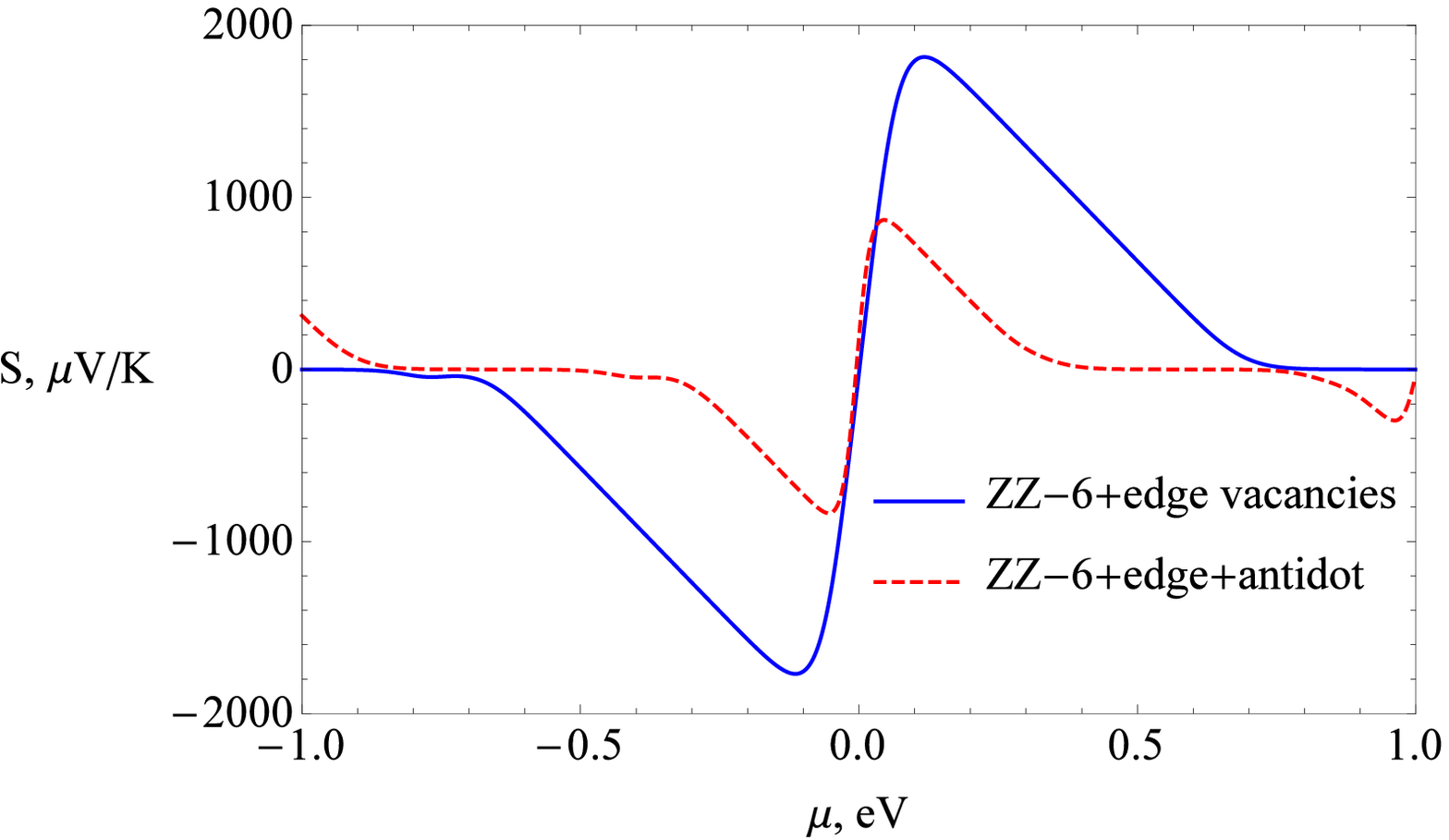} 
\includegraphics{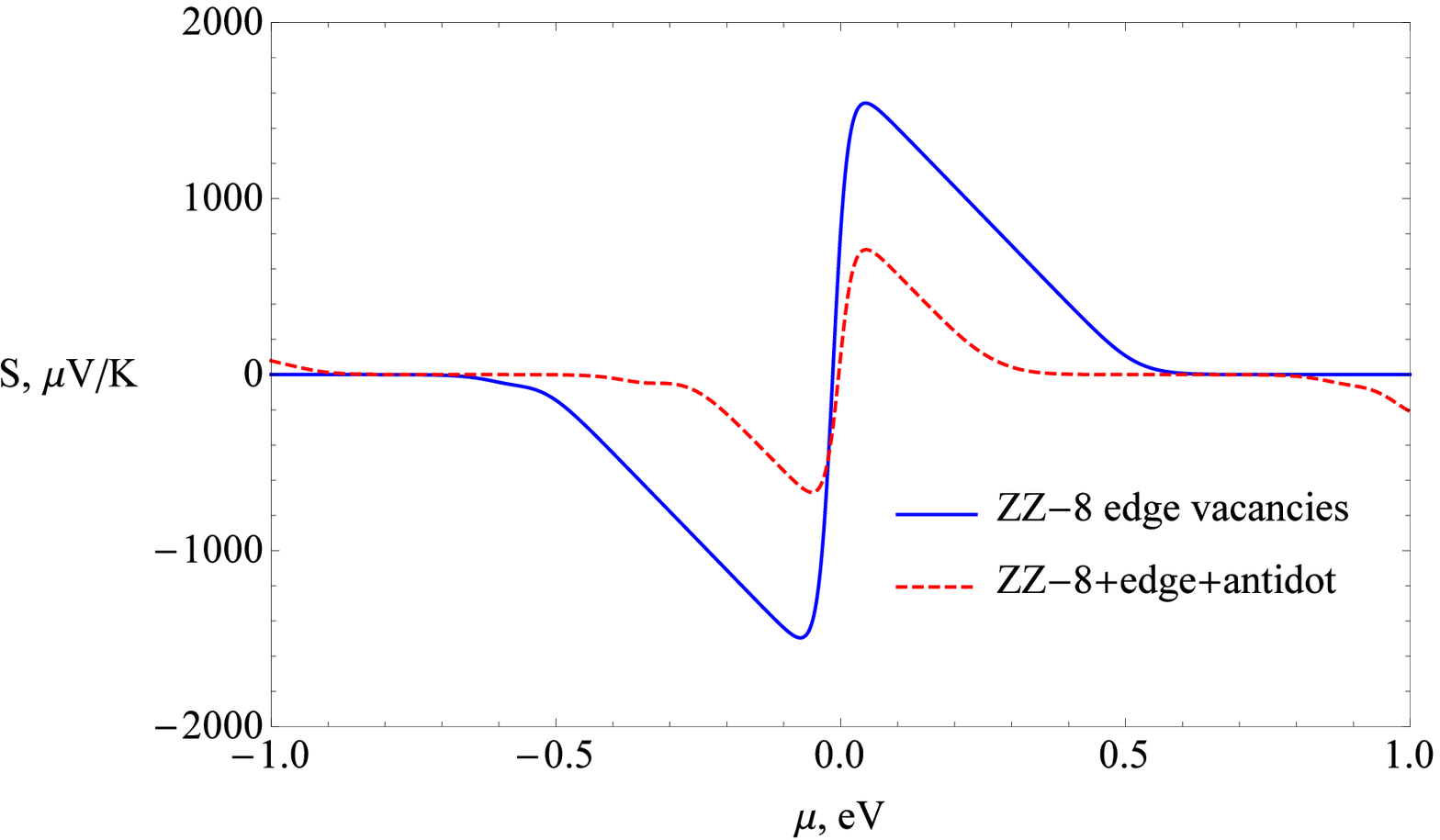} 
}
\caption{The Seebeck coefficient at T=300 K for the 6 and 8 atoms wide ZGNR  with edge vacancies (solid line) and additional periodic antidot (dashed line).}
\label{fig-Zee}
\end{figure*}
The Seebeck coefficient at the room temperature for the structures with and without periodic antidot is shown in Fig.~\ref{fig-Zee}. Since the value of $\Delta E$ is bigger for the case of antidot (see table \ref{t1}), the introduction of antidot decreases S. One should also note that the peak values of S, both with and without antidots, at the room temperature are sufficiently greater than those in other  nanostructures with high ZT discussed elsewhere~\cite{sevinci,mazzu,cunni}. This directly follows from the specific behavior of the electron transmission function, having the marked step height ($T_{max}=2$) and low $\Delta E$.
\begin{figure*}
\resizebox{0.9\textwidth}{!}{%
\includegraphics{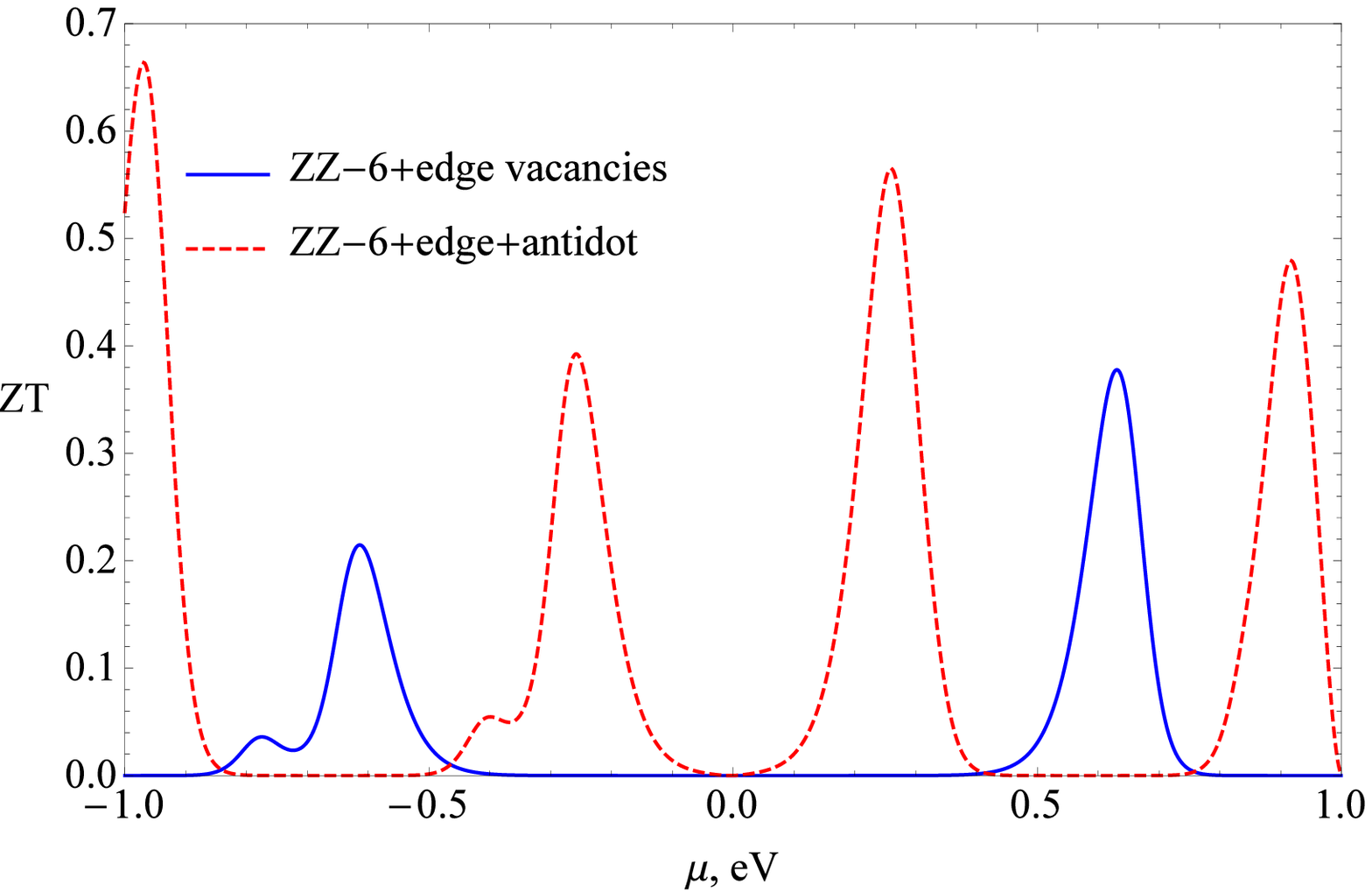} 
\includegraphics{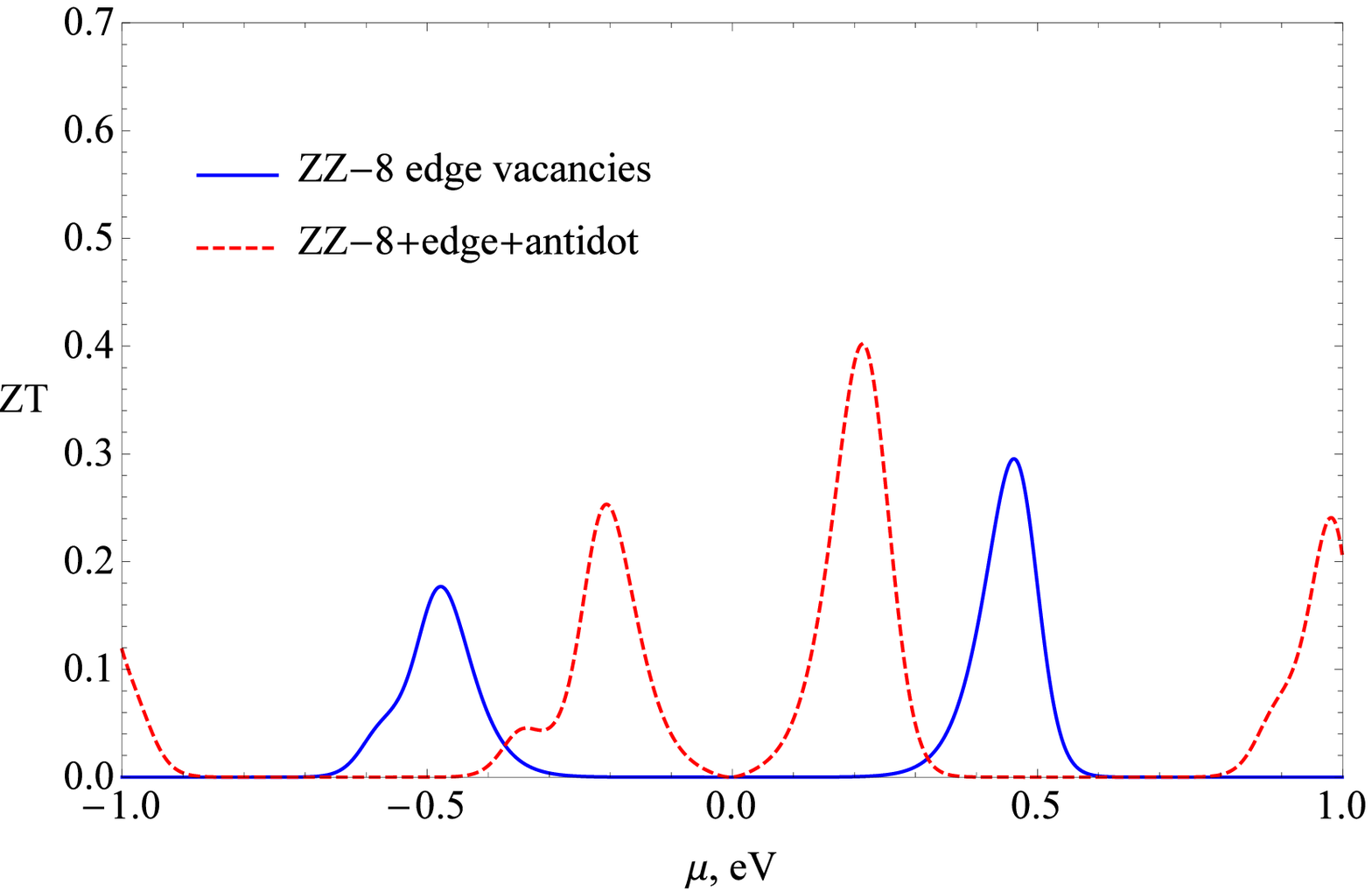} 
}
\caption{The thermoelectric figure of merit at T=300 K 6 and 8 atoms wide ZGNR  with edge vacancies (solid line) and additional periodic antidot (dashed line).}
\label{fig-ZT}
\end{figure*}
Fig.~\ref{fig-ZT} shows the thermoelectric figure of merit calculated at room temperature. As is seen, the highest peak values of ZT, corresponding to the band edges, are 0.35 and 0.3 for structures with the width of 6 and 8 atoms, respectively, with edge vacancies, and 0.55 and 0.4 for the case of the combination of vacancies with periodic antidot. Additional peaks of ZT at $\mu\approx \pm 1$ eV with the values up to 0.25 for the wide structure and 0.65 for the narrow one are generated by the secondary transmission function steps, which appear due to the introduction of periodic antidots.  This behavior is the result of two factors: reduced thermal conductance and the degenerate character of the band edge states. The thermal conductance (shown in Table~\ref{t1}) is increased for the wider structure and decreases due to the presence of antidots.

It should be mentioned that the 4th nearest neighbor method typically overestimates the thermal conductance (see, e.g. Ref.~\cite{huang}), so that the actual ZT are expected to be above the calculated values. In the conduction band, $\Delta E$  is found to have a lower value for the wider structure. The presence of periodic antidot increases the sharpness of the electron transmission function on the band edge for both widths, which means a slightly negative influence on the electronic part of the figure of merit. For the antidot case, the combination of reduced thermal conductance with smoothed electron transmission on the band edge results in the increased ZT for ZGNRs of both widths. 
\begin{table*}
\caption{The phonon thermal conductance $\kappa_{ph}$ at T=300 K, distance between energy levels $\Delta E$ at the edge of conduction (+) and valence (-) bands, and the band gap $E_G$ for the ZGNRs with edge vacancies and antidots.}
\label{t1}
\begin{tabular}{l|l|l|l|l}
Structure & ZZ-6+vac.&ZZ-6+vac.+antidot& ZZ-8+vac.&ZZ-8+vac.+antidot\\
\hline
$\kappa_{ph}$, nW/K & 0.9917  & 0.5282  & 1.376 & 0.8556 \\ \hline
$\Delta E_{+}$, eV & 0.016 & 0.042 &  $<$0.001 & 0.027 \\ \hline
$\Delta E_{-}$, eV & 0.135 & 0.118 & 0.081 & 0.112 \\ \hline
$E_G$, eV & 1.3 & 0.577 & 0.999 & 0.478 \\
\end{tabular}
\end{table*}

\section{Conclusion}

We have investigated the thermoelectric properties of zigzag graphene nanoribbons, 6 and 8 atoms wide, using the Green's function method with the tight-binding approximation for electronic Hamiltonian and 4th nearest neighbour approximation for the phonon dynamical matrix. The electron and phonon transmission functions, Seebeck coefficient, total thermal conductance and thermoelectric figure of merit were calculated. 

For the unique case of periodic vacancies on both ZGNR edges with the period equal to three elementary translations, we have found the appearance of nearly degenerate electronic states near the conduction and valence band edges. These states ensure the presence of the sharp step-like increase of electron transmission on the short energy interval, which varies from 0.1 to 6 $k_B T$ at the room temperature. The presence of the next-to-nearest terms in the electronic Hamiltonian ensures the asymmetry of the electron transmission function, decreasing $\Delta E$ for one band edge and increasing it for another. This behavior results in the increase of Seebeck coefficient and the figure of merit for the chemical potential near the band edges, where higher values of S and ZT correspond to the lover $\Delta E$. To decrease the phonon thermal conductance and increase the value of figure of merit even further, we add the periodic antidot lattice to the structure with the period equal to 6 elementary translations. The combination of antidot lattice and edge vacancies generally preserves the degeneracy of the states on the valence and conduction band edges and decreases the band gap. For such structures, we found the peak value of figure of merit for 6 atom wide ZGNR to be 0.55, and for 8 atom wide ZGNR to be 0.45. The decreased thermal conductance due to the periodic antidot was established to be the crucial factor leading to the increase of ZT. One should also note that the presence of antidot leads to the decrease of the chemical potential, that is necessary to achieve the maximum thermoelectric efficiency.  The top values of the Seebeck coefficient and  thermoelectric efficiency were also found to be larger for the 6 atom wide ZGNR, both with edge vacancies and vacancy-antidot combination, due to the reduced phonon conductance in the wider structure. 
\begin{acknowledgments}
We would like to thank Dr. V. L. Katkov for fruitful discussions.
\end{acknowledgments}

\end{document}